\documentclass[manuscript, screen, anonymous=false, review=false]{acmart}
\AtBeginDocument{%
  }

\setcopyright{acmlicensed}
\copyrightyear{2018}
\acmYear{2018}
\acmDOI{XXXXXXX.XXXXXXX}
\acmConference[Conference acronym 'XX]{Make sure to enter the correct
  conference title from your rights confirmation email}{June 03--05,
  2018}{Woodstock, NY}
\acmISBN{978-1-4503-XXXX-X/2018/06}

\begin{document}

\title{Perception of Deepfakes among Bangladeshi Women}


\author{Sharifa Sultana}
\affiliation{%
  \institution{University of Illinois Urbana-Champaign}
  \country{USA}}
\email{sharifas@illinois.edu}

\author{Pratyasha Saha}
\affiliation{%
  \institution{University of Illinois Urbana Champaign}
  \country{USA}}
\email{saha19@illinois.edu}

\author{Nadira Nowsher}
\affiliation{%
  \institution{Eden Mohilla College, Dhaka}
  \country{Bangladesh}}
\email{nowshernadira@gmail.com}

\author{Sumaia Arefin Ritu}
\affiliation{%
  \institution{Brac University}
  \country{Bangladesh}}
\email{sumaiaritu550@gmail.com}

\author{Zinnat Sultana}
\affiliation{%
  \institution{S.M.R. Law College, Jessore}
  \country{Bangladesh}}
\email{zinnat1409@gmail.com}

\author{Syed Ishtiaque Ahmed}
\affiliation{%
  \institution{University of Toronto}
  \country{Canada}}
\email{ishtiaque@cs.toronto.edu}

\author{S M Taiabul Haque}
\affiliation{%
  \institution{Brac University}
  \country{Bangladesh}}
\email{eresh03@gmail.com}

\renewcommand{\shortauthors}{Sultana et al.}

\begin{abstract}
 As deepfake technology becomes more accessible, concerns about its misuse and societal impact are escalating, particularly in regions like the Global South where digital literacy and regulatory measures are often limited. While previous research has explored deepfakes in contexts such as detection and media manipulation, there is a noticeable gap in understanding how individuals in these regions perceive and interact with deepfake media. This study addresses this gap by investigating how Bangladeshi women perceive deepfakes and the socio-cultural factors influencing their awareness, concerns, and responses to this technology. Drawing on 15 semi-structured interviews, we uncover how cultural values, gendered norms, trust in institutions, and the prevalence of digital harassment shape their perceptions and coping mechanisms. Through this research, we aim to advance existing scholarship in HCI by offering insights into the design of culturally sensitive interventions, educational initiatives, and policy frameworks to address the challenges posed by deepfakes in the Global South.
\end{abstract}



\maketitle

\section{Introduction and Background}
Deepfake technologies such as AI-generated or AI-manipulated synthetic media across images, video, audio, and text, have rapidly shifted from a technical novelty to a widely accessible tool for producing persuasive, realistic-looking content \cite{mirsky2021creation}. While deepfakes offer creative possibilities in entertainment, education, and media production, they also pose serious ethical, societal, and legal risks \cite{chesney2019deepfakes, citron2014criminalizing}. By enabling the non-consensual manipulation of a person’s likeness, deepfakes blur the boundary between reality and fabrication, undermining trust, privacy, and reputational integrity \cite{chesney2019deepfakes}. These harms are especially acute when deepfake content circulates at scale through social media platforms, where visibility, virality, and moral judgment are tightly coupled \cite{vaccari2020deepfakes}.

Although deepfakes have become a global concern, their impacts are unevenly distributed across social and cultural contexts. In patriarchal societies such as Bangladesh, deepfakes are increasingly used as tools for harassment, defamation, and sexualized abuse, disproportionately targeting women \cite{citron2014criminalizing, flynn2019image, henry2020image}. A single manipulated image or video can trigger rapid social circulation, family intervention, institutional exclusion, and long-term reputational damage. These harms are compounded by limited digital literacy, weak platform accountability, cultural stigma surrounding sexuality, and inadequate legal response mechanisms \cite{irani2010postcolonial, sambasivan2021re, sambasivan2021seeing}. Yet, despite the urgency of these risks, empirical research examining how deepfakes are perceived, experienced, and navigated in non-Western contexts remains limited \cite{irani2010postcolonial, sambasivan2021re, sambasivan2021seeing, dourish2012ubicomp}.

Existing scholarship on deepfakes has largely focused on technical detection, model robustness, and legal or regulatory implications \cite{hsu2020deep, chang2020deepfake, patel2023improved, raza2022novel, uma2024securing, gandhi2020adversarial, song2023robustness, meskys2020regulating, kugler2021deepfake, tuysuz2023analyzing, mania2024legal}. While some studies examine societal effects—such as misinformation, political manipulation, and declining trust in media—empirical work on everyday perceptions and lived experiences is still nascent \cite{park2020global, moy2000media, ognyanova2020misinformation}. Research in communication and psychology suggests that deepfakes often generate uncertainty rather than outright deception, but this uncertainty itself can erode trust and intensify harm \cite{vaccari2020deepfakes, hameleers2024they}. Within HCI, prior work has primarily explored detection tools or training interventions, with relatively little attention to how deepfake harms unfold in culturally specific settings where verification, reporting, and repair are socially constrained \cite{gamage2022deepfakes, danry2022ai, kaate2023users}. As a result, we lack human-centered accounts of how people assess authenticity, manage risk, and decide whether seeking help is possible or safe.

To address this gap, this ongoing research examines how Bangladeshi women perceive deepfake technologies and navigate the social, cultural, and institutional conditions surrounding them. We ask: (1) How do Bangladeshi women understand and emotionally respond to deepfakes? (2) How do gender norms, family dynamics, and cultural taboos shape the harms deepfakes produce? and (3) What strategies do women use to verify content, manage privacy, and evaluate legal recourse? We conducted 15 semi-structured interviews with women aged 18–45 from diverse educational and professional backgrounds. Interviews were conducted in Bengali, transcribed, and analyzed using thematic analysis to identify recurring patterns in perception, fear, and coping.

Our findings show that participants rarely viewed deepfakes as entertainment; instead, they described them as inherently dangerous and easily weaponized. The primary fear was not manipulation alone, but how quickly deepfake content could circulate and be treated as credible evidence. Gendered reputational norms intensified harm, as women were more likely to be blamed, morally judged, and socially sanctioned—even when content was known to be fake. Participants also reported limited awareness and weak verification capacity, relying on informal heuristics rather than reliable tools. Legal recourse was widely perceived as inaccessible or risky, leading many to prioritize silence over reporting. Together, these findings highlight how deepfake harms are produced through sociotechnical systems rather than technology alone. This paper contributes empirical evidence from an underexplored context, advancing human-centered understandings of deepfake risk and informing culturally sensitive design and policy interventions as part of an ongoing research agenda.

\section{Methodology}

Deepfake generation technologies continue to advance in realism and accessibility, increasing both their reach and potential for harm. To examine how these risks are perceived and experienced, we adopted a qualitative research approach (e.g., interviews) centered on the lived experiences of Bangladeshi women \cite{kvale2009interviews}. Interviews are particularly appropriate for studying deepfake-related harm because such experiences are socially embedded, emotionally charged, and shaped by cultural norms that are difficult to capture through quantitative measures \cite{kvale2009interviews}. Our primary data collection method consisted of semi-structured interviews with 15 women aged between 18 and 45 from diverse backgrounds across Bangladesh. All participants provided informed consent prior to participation.

\subsection{Participant Recruitment and Interview Method}
Participants were recruited using purposive and snowball sampling to ensure diversity in age, education, occupation, and familiarity with digital technologies, including students, professionals, NGO workers, and homemakers. Interviews were conducted one-on-one using a semi-structured format to balance consistency with flexibility, allowing participants to share personal experiences and reflections while covering core topics such as awareness of deepfakes, perceived risks, emotional and social impacts, verification practices, legal recourse, and privacy strategies. Each interview lasted approximately 30–45 minutes and was conducted in Bengali to ensure participants could comfortably express themselves. All interviews were audio-recorded with participants’ consent and later transcribed verbatim.

\subsection{Data Collection}
Data were collected over several weeks through in-person or remote interviews, depending on participants’ availability and comfort. Before each interview, participants were informed about the study’s purpose, the voluntary nature of participation, and confidentiality measures, and pseudonyms were assigned to protect identities. The interview guide was iteratively refined during early data collection to better capture emerging issues such as fear of retaliation, family reactions, and the perceived irreversibility of deepfake harm. In addition to audio recordings, the interviewer maintained brief field notes to document contextual details and reflections that informed later analysis.

\subsection{Data Analysis}
We analyzed the data using reflexive thematic analysis \cite{braun2006using}, which enabled us to identify recurring patterns while remaining attentive to sociocultural context. Transcripts were manually coded using open coding techniques, followed by iterative grouping of codes into higher-level themes through constant comparison across participants. The research team discussed and refined themes to ensure analytic rigor and coherence, with particular attention to how gender norms, cultural expectations, and structural constraints shaped participants’ experiences. Reflexive practices were employed throughout the analysis to account for the researcher's positionality and to ground interpretations in participants’ narratives.

\section{Findings}

We begin with a vignette that captures the lived experience of deepfake abuse faced by Bangladeshi women. This account, from Riya (pseudonym), reflects patterns echoed across our interviews.\\

\textit{Riya (23) is a university student in Rajshahi and actively engaged in several voluntary student organizations. Due to her outspoken presence, she gained visibility among her peers. In her senior year, some peers circulated sexually explicit images in student networks where her face was manipulated. The content quickly spread through social media, including among friends and relatives. Riya was pressured to resign from all student organizations she was affiliated with. When her parents found out, her mother told her to quit her studies and leave campus. Riya wanted to seek legal help, but believed the police could not identify perpetrators and feared that filing a complaint would attract media attention and intensify the harm. She also worried that legal action might provoke further retaliation.}\\

Riya’s story illustrates how deepfakes generate cascading harms that extend beyond digital spaces into educational, familial, and institutional domains. Similar dynamics appeared across participants’ accounts, shaping how they understood, feared, and responded to deepfake technologies.

\subsection{Deepfake? It’s never fun!}

Although deepfakes are often framed as playful or entertaining technologies, participants overwhelmingly rejected this characterization. They described deepfakes as inherently risky because manipulated content can escape its original context and later be repurposed for harassment, coercion, or revenge.

P7 framed deepfakes as a mechanism of targeting rather than amusement:
\begin{quote}
Nowadays, people use deepfakes to poke fun at or bully someone, to harass them, or even to create political controversy. It’s almost like a weapon, a lot of individuals are targeted this way. For me, the concept of deepfakes has taken on a negative connotation. This can never be fun!
\end{quote}

Participants’ fears were grounded in lived experience. P14 described how intimate content created within a relationship was later weaponized:
\begin{quote}
In my previous relationship, my boyfriend and I used to make fun intimate videos of us using an app called Luma. They were like kissing videos, hugging videos, etc. Unfortunately that relationship didn’t work out. (...) After we broke up and I got married, he sent those intimate videos to my husband. It was so embarrassing and humiliating! I was terrified it might lead to a divorce. I had just gotten married, and it was deeply traumatic for both me and my husband. I felt completely helpless. I never imagined he would use something so personal out of revenge. This is insane!
\end{quote}

Across accounts, participants emphasized the irreversibility of deepfakes: content created “for fun” could later resurface in moments of social vulnerability, making any participation feel unsafe and difficult to undo.

\subsection{Understanding the Fear: Deepfakes, Gender, and Cultural Trauma}

Participants situated their fear of deepfakes within Bangladesh’s patriarchal social order, where women’s reputations are closely tied to moral judgment and sexual respectability. When sexually explicit deepfake content circulates, women are often presumed guilty rather than harmed.

P7 articulated this gendered asymmetry:
\begin{quote}
Women have the pressure to maintain a very good reputation in our society rather than men. If a deepfake content of a girl gets viral, instantly everyone thinks that the girl must be `bad'. They will pick it up, share it among themselves, and keep bringing it up in every situation. But if it's a boy, they will say, ``Oh he is just a boy, let it go!''. It's not like this does not happen to the men, but when it happens with a girl it leaves a hundred times worse impact on her life.
\end{quote}

Participants also described how cultural taboos around sexuality restrict women’s ability to explain or defend themselves, especially within families. P10 explained:
\begin{quote}
If something like this ever happened to me, how would I even talk to my parents about it? In our culture, topics like these are so taboo, we’re not supposed to speak about them with our parents or elders. It feels inappropriate, even shameful (...) and if I can’t talk about it, how would I even begin to explain what happened? How would they ever understand?
\end{quote}

Together, these accounts show that deepfake harm is amplified by cultural norms that presume female culpability and constrain women’s ability to contest manipulated evidence, transforming digital falsification into enduring social and emotional trauma.

\subsection{Limited Awareness and Literacy about Deepfakes}

Participants consistently reported low awareness of how deepfakes are created and how authenticity might be assessed. This lack of literacy, they noted, leads to widespread belief in manipulated content and delays in responding to harm.

P2 observed this gap across regions:
\begin{quote}
As an NGO worker, I’ve traveled across various regions of Bangladesh and interacted with people from diverse backgrounds. The practice of manipulating images or videos to target and harm certain individuals or groups has existed for years. But most people don’t even know that deepfake content can exist, or that something like this can be made at all.
\end{quote}

Even participants who were aware of deepfakes described limited capacity to verify content. P13 said:
\begin{quote}
A lot of images I see online are personally or emotionally triggering, but I don't understand if they are edited or not. I actually don't always have the head space to check those. And I don't know about any platforms or tools to detect these.
\end{quote}

As a result, participants relied on fragile heuristics and social cues rather than formal tools, allowing deepfakes to circulate faster than individuals could verify, contest, or contextualize them.

\subsection{Weak Legal Infrastructure}

Despite severe consequences, participants expressed little confidence in legal remedies. Most were unaware of applicable laws and believed that law enforcement lacked preparedness and sensitivity.

P16 summarized this perception:
\begin{quote}
No, I don’t think they are prepared at all. Deepfake content is being shared openly on platforms like Facebook. No matter what the purpose is, there’s no regulation around it. And if there’s no law, then how are they supposed to handle these cases?
\end{quote}

Participants also feared judgment and blame from authorities. One participant explained:
\begin{quote}
If you go to the police station and say that your fake video has gone viral, they look at you like it’s your fault. They don’t understand how these things work. Sometimes they even say things like, “Why did you share your photos on social media in the first place?”
\end{quote}

Anticipated disbelief, moral scrutiny, and procedural burden led many participants to view silence as safer than reporting, reinforcing underreporting and prolonging harm.

\subsection{Privacy Strategies to Combat Deepfake-related Harm}

Participants described privacy management as necessary but insufficient. While many restricted audiences or avoided posting identifiable images, they emphasized that complete protection is impossible once content exists online.

P4 articulated this sense of inevitability:
\begin{quote}
I am breaching my own privacy with my own hands. I have my own website. My photo, my information, these can be seen by everyone out there. Even if I lock my social media profile, I see my photos available in Google search. Anyone can do anything with that!
\end{quote}

Participants also described trade-offs between safety and self-expression. P17 said:
\begin{quote}
I frequently travel to different countries, so I post a lot of photos on my socials. I love sharing those with my followers. I know about deepfakes, but I cannot stop doing what I love. My socials are like my digital diary to me.
\end{quote}

These accounts show how privacy strategies are constrained by technical visibility and social expectations, positioning women to continually negotiate between participation and protection.

\section{Discussion}
Our findings show that deepfake harm for Bangladeshi women is not primarily a problem of technical deception, but one of sociocultural interpretation and power \cite{selbst2019fairness, raji2020closing}. Participants’ fear centered less on whether content was real and more on how quickly manipulated media could be accepted as credible evidence, triggering moral judgment, institutional exclusion, and family intervention. In patriarchal contexts where women’s reputations are tightly policed, deepfakes act as credibility accelerators: even when known to be fake, they retain the power to silence, shame, and displace women from public and professional spaces. This reframes deepfakes from a detection-centric challenge to a sociotechnical one, where harm emerges through the interaction of platform virality, gendered moral norms, and weak accountability infrastructures \cite{selbst2019fairness}. For HCI, this underscores the limits of accuracy-focused solutions and calls for designs that support contestability, social repair, and harm mitigation rather than verification alone.

From a Feminist HCI and Responsible AI perspective, our findings show how deepfake technologies intensify existing gendered asymmetries rather than operating as neutral tools \cite{bardzell2010feminist}. Women’s constrained ability to safely explain, contest, or report abuse reflects structural limits on voice and legitimacy. Likewise, Responsible AI approaches that focus narrowly on transparency or fairness overlook harms produced downstream through social interpretation and institutional failure \cite{raji2020closing, selbst2019fairness}. Participants’ reliance on informal verification heuristics and avoidance of legal systems reveal how accountability gaps shift risk management onto individuals—particularly women—who already bear disproportionate social and emotional costs. These insights emphasize the need for human-centered AI that accounts for cultural stigma, emotional labor, and asymmetric exposure to harm, especially in Global South contexts.

\subsection{Design Implications for Gender-Sensitive Deepfake Mitigation}
Our findings suggest several design implications for HCI and Responsible AI practitioners. First, mitigation should move beyond technical detection toward socially legible mechanisms of contestation, such as context-aware labels, rapid takedown signals, or third-party attestations that allow victims to challenge content without public self-disclosure. Second, platform reporting and moderation workflows must be trauma-aware and gender-sensitive, reducing secondary victimization and the burden of proof placed on victims. Third, awareness and literacy interventions should be community-facing and culturally grounded, acknowledging taboos around sexuality and family disclosure. Finally, privacy-by-design approaches must recognize the inevitability of online visibility and support harm-reduction strategies that do not require women to withdraw from digital participation altogether.

\subsection{Limitations and Future Work}
This study has limitations. Our sample size is modest and includes women with some level of internet access, which may not capture experiences in more rural or digitally excluded settings. As an interview-based study, our findings reflect reported experiences rather than direct observation of platform governance or legal processes, and we do not include perspectives from platform operators, law enforcement, or policymakers.

As ongoing work, we plan to expand participant diversity across geographic and socioeconomic contexts and to engage institutional stakeholders to better understand structural constraints. We also aim to explore participatory and co-designed interventions that support culturally grounded reporting, contestation, and repair mechanisms for deepfake-related harm.

\bibliographystyle{ACM-Reference-Format}
\bibliography{sample-base}

\end{document}